\documentclass[%
superscriptaddress,
twocolumn,
showpacs,preprintnumbers,
 amsmath,amssymb,
 aps,
prb,
]{revtex4}
\usepackage{graphicx}
\usepackage{dcolumn}
\usepackage{bm}
\usepackage{hyperref}
\usepackage{multirow}
\usepackage[usenames,dvipsnames]{color}
\begin{document}
\newcommand{\bt}[1]{\textcolor{blue}{#1}}
\preprint{}

\title{Anderson lattice with explicit Kondo coupling: general features and the field-induced suppression of heavy-fermion state in ferromagnetic phase}

\author{Olga Howczak}
\email{olga.howczak@uj.edu.pl}   
\affiliation{Marian Smoluchowski Institute of Physics, Jagiellonian University, 30-059 Krak\'ow, Poland}

\author{Jozef Spa\l ek}
\email{ufspalek@if.uj.edu.pl} 
\affiliation{Marian Smoluchowski Institute of Physics, Jagiellonian University, 30-059 Krak\'ow, Poland} 
\affiliation{AGH University of Science and Technology, Faculty of Physics and Applied Computer Science, 30-059 Krak\'ow, Poland.}
\date{\today}

\date{\today}

\begin{abstract}
We apply the extended (statistically-consistent, SGA) Gutzwiller-type  approach to the periodic Anderson model (PAM) in an applied
magnetic field and in the strong correlation limit. The finite-U corrections are included systematically by transforming PAM into the form with Kondo-type interaction and residual hybridization, appearing both at the same time. This effective Hamiltonian represents the essence of \textit{Anderson-Kondo lattice model}. We show that in  ferromagnetic phases the low-energy single-particle states are strongly affected by the presence of the applied magnetic field. We also find that for large values of
hybridization strength the system enters the so-called \textit{locked heavy fermion state}.
In this state the chemical potential lies in the majority-spin hybridization gap and as a consequence, the
system evolution is insensitive to further increase of the applied field. However, for a sufficiently strong
magnetic field, the system transforms from  the locked state to the fully spin-polarized phase. This
is accompanied by a metamagnetic transition, as well as by drastic reduction of the effective mass of
quasiparticles. In particular, we observe a reduction of effective mass enhancement in the majority-spin subband by as
much as 20\% in the fully polarized state. The findings are consistent with experimental results for Ce$_x$La$_{1-x}$B$_6$ compounds. The mass enhancement for the spin-minority electrons may also diminish with the increasing field,
unlike for the  quasiparticles states  in a single narrow band in the same limit of strong correlations.

\end{abstract}

\maketitle


\section{Introduction}

Heavy fermion (HF) metals are customarily regarded  as a quantum liquid composed of strongly correlated (quasi-atomic) and heavy f-electron 
quasiparticle states resulting from hybridization with conduction (band) 
electrons.  The starting f-electrons are associated with the presence of rare-earth or actinide ions (ions with valency close to integer valence, e.g. Ce$^{+3-\delta}$, with $\delta \ll 1$). In other words, the f-f on-site Coulomb repulsion is strong and the f-electrons behave often in a solid as localized (or almost localized) magnetic moments even in the presence of the interband hybridization. Moreover, those heavy hybrid f-electrons are coupled to the conduction (d-s) electrons via antiferromagnetic Kondo interaction \cite{RevModPhys.79.1015}. Also, many of the cerium or uranium HF compounds order antiferromagnetically. Nonetheless, there is also a constantly growing group of heavy fermion compounds which show a ferromagnetic   ordering, e.g. URu$_{2-x}$Re$_x$Si$_2$ for $0.3 < x < 1.0$  \cite{PhysRevLett.94.046401},   UIr$_{2}$Zn$_{20}$ \cite{PhysRevB.74.155118}, 
CeSi$_{1.81}$ \cite{PhysRevB.73.214413},YbRh$_{2}$Si$_{2}$ \cite{PhysRevLett.94.226402}. Therefore, it is important to understand the complexity
of their phase diagram within a single theoretical framework such as Anderson- \cite{PhysRev.124.41} or Kondo-lattice models \cite{PTP.32.37}. 
In this respect, we use here a mixed Anderson-Kondo lattice picture derived from PAM in a systematic manner.

One of other characteristic properties of the heavy fermion
compounds is the metamagnetic behavior or a sudden magnetization increase 
  (\textit{metamagnetism}) at a critical value of the applied magnetic field, observed in HF compounds such as CeRu$_2$Si$_2$ \cite{springerlink:10.1007/BF00710351}, CeCoGe$_3$ \cite{PhysRevB.47.11839},
 UPd$_2$Al$_3$, URu$_2$Si$_2$, and UPt$_3$ \cite{Sugiyama2000244}.  This feature of HF electron systems makes the applied field a useful  tool of probing the  strong-correlation behavior. For example, a sufficiently strong applied field may lead to a destruction of the HF state of those strongly correlated fermions and, as a consequence, force a localization (freezing) of the heavy, predominantly f-electron, quantum liquid. 
This type of transition signalled by a metamagnetic transition (MMT), was the subject of extensive studies during the past decade, but still,
because of the rich variety and complicated structure of their phase diagrams, the universal nature of the
MMT is hard to establish. One of the puzzles is that the metamagnetic behavior can take place starting from the ferromagnetic state.

In this paper we present a scenario in which the heavy itinerant f-electron quasiparticles localize at
MMT, with a concomitant suppression of  the HF state at that critical field. To account for this interplay between the localized- and the itinerant-type behavior we start from the periodic Anderson model (PAM) in an  external magnetic field included via the Zeeman term.
We also use the large, but finite $U$ version of PAM formulated before \cite{Gopalan&JS, PhysRevB.38.208}, in which both
residual hybridization and the Kondo-type coupling are included in a systematic manner. We have used this model before
\cite{PhysRevB.82.054509, PhysRevB.49.1454}, which can be called the Anderson-Kondo lattice model, to discuss the so-called hybrid (Kondo-type) pairing in heavy-fermion systems. Additionally we utilize the so-called \textit{statistically-consitent} extension of the Gutzwiller-type approach \cite{PhysRevB.81.073108, 2010arXiv1008.0021J}. Such formulation allows for a consistent mean-field discussion of correlated states, here in the applied magnetic field. 
 With this paper we are starting a series with a detailed analysis of various magnetic, as well as of the magnetic superconducting states. 

The paper is organized as follows. In Sec. \ref{model} we summarize our theoretical model.  In Sec. \ref{results} we discuss the so-called statistically consistent mean-field approach (SGA) and provide  numerical results, as well as apply them to the analysis of Ce$_x$La$_{1-x}$B$_6$ properties. We summarize our results and provide a brief outlook in Sec. \ref{summary}. The Appendices \ref{A2}-\ref{A1} provide some details of our analytic part of the approach.

\section{Model and statistically-consitent mean field theory}
\label{model}
\subsection{Anderson-Kondo lattice Hamiltonian}
In modeling the fluctuating-valence and heavy fermion systems one usually starts from the periodic Anderson model (PAM) \cite{uwaga},
which in the real space (Wannier) representation reads 
\begin{eqnarray}
\mathcal{H} &= &\sum_{mn\sigma}t_{mn} c^{\dagger}_{m\sigma} c_{n\sigma} + \epsilon_f \sum_{i\sigma} \hat{N}_{i\sigma} + U \sum_{i\sigma} \hat{N}_{i\uparrow} 
\hat{N}_{i\downarrow} \nonumber\\
&+& \sum_{im\sigma} \left( V_{im} f^{\dagger}_{i\sigma} c_{m\sigma} + H.c. \right) - \mu \left( \sum_{i\sigma} \hat{N}_{i\sigma} + \sum_{m\sigma} \hat{n}_{m\sigma}\right)\nonumber\\
& -& \frac{1}{2} g_f\mu_B H \sum_{i\sigma} \sigma \hat{N}_{i\sigma} - \frac{1}{2} g_c\mu_B H \sum_{m\sigma} \sigma \hat{n}_{m\sigma},
\label{eq:1} 
\end{eqnarray}
with $\hat{N}_{i\sigma} \equiv f^{\dagger}_{i\sigma}f_{i\sigma}$ being the number of originally atomic (f) electrons and $\hat{n}_{m\sigma} \equiv c^{\dagger}_{m\sigma}c_{m\sigma}$ the number of conducting (c) electrons. The consecutive terms are: the first represents the hoping (band) energy of c electrons, the second - bare f atomic level position (with respect to the atomic  level of c electrons, $\epsilon_f \equiv \epsilon_f-\epsilon_c$), the third - interatomic 
Coulomb interaction among the f-electrons (the Hubbard term), the fourth the f-c hybridization (with the amplitude $V_{im}$, taken in the subsequent analysis in the intraatomic form, $V_{im} = \delta_{im}V$), and the next term represents the subtraction of the chemical potential part, as we will work in the grand canonical scheme. The last two terms are the Zeeman energy in the applied magnetic field H of f and c electrons, respectively. 

In the context of heavy fermion system this Hamiltonian has usually been taken in the $U= \infty$ limit, as this parameter represents the 
highest energy scale in the system. Typically for Ce systems: $U=5-6eV$, $\epsilon_f=-1\div-2 eV$, $V\lesssim -0.5eV$, $W = 2z|t| \sim 2 eV$ \cite{PhysRevLett.103.096403}. While $|V|/U\ll 1$, $U$ is definitely finite and therefore, the finite-U corrections should be taken into account, as e.g. the residual Kondo exchange interaction of the magnitude $J_K = 2V^2/(\epsilon_f+U)\sim 0.1eV$ arises \cite{PhysRevB.38.208} and will influence  the nature of the magnetic ground state in a decisive manner. However, to account for the Kondo interaction systematically, as well as to allow for itineracy of f-electrons at the same time, a direct application of either the Schrieffer-Wolff \cite{PhysRev.149.491} transformation or by starting from the periodic Kondo model \cite{PTP.32.37} (i.e. neglecting explicitly the hybridization term apart from the presence in the expression $\sim J_K$), may not be realistic. At least, one has to try to reach such Kondo-lattice limit in a systematic manner, by e.g. including explicitly both the nonzero renormalized hybridization and the nonvanishing Kondo coupling in the large-U limit at the same time.
\begin{figure}
  \centering
 \includegraphics[width=0.5\textwidth]{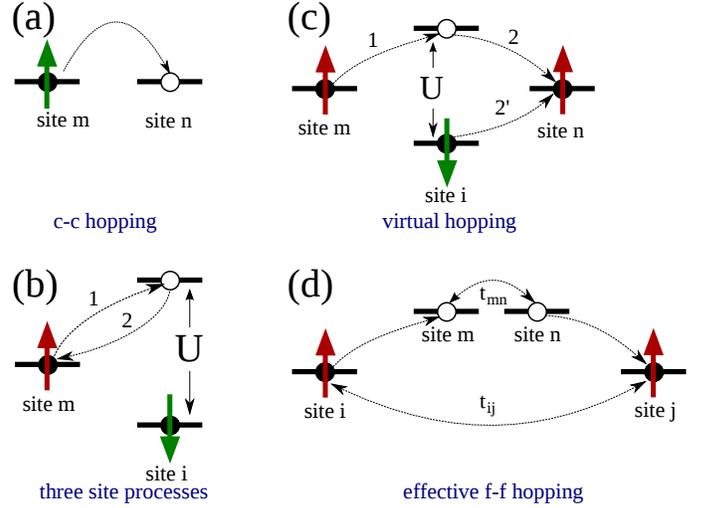}                
  \caption{(Color online) Schematic representation of the first (a) and second-order (b-d) hopping processes. The hopping label $2$ and $2'$ in (c) are alternative processes. The process (b) and (c) lead to real-space pairing, whereas the three-step process (d) leads to itineracy of originally atomic f electrons with effective hopping $t_{ij}$. All the processes contribute to the dynamics of heavy quasiparticles with renormalized characteristics. The virtual process (b) leads to the Kondo interaction, whereas the process (c) introduces hybrid-pair hopping. The effective f-f hopping (d) results from residual hybridization in the narrow-band limit (for detailed discussion of this last picture see the next Subsection \ref{B}).}
\label{fig:1}
\end{figure}

For this purpose, in direct analogy to the corresponding projection set originally for the Hubbard model \cite{APPSpalek}, we divide
the f-c hybridization into the following two parts according to the prescription

\begin{eqnarray}
f^{\dagger}_{i\sigma} c_{m\sigma}& \equiv& (1-\hat{N}_{i\bar{\sigma}}+\hat{N}_{i\bar{\sigma}} ) f^{\dagger}_{i\sigma} c_{m\sigma}\nonumber\\
&=& (1-\hat{N}_{i\bar{\sigma}})f^{\dagger}_{i\sigma} c_{m\sigma}  + \hat{N}_{i\bar{\sigma}} f^{\dagger}_{i\sigma} c_{m\sigma}.
\label{eq:2}
\end{eqnarray}          
The first term represents the projected hybridization of the states  $|f_{i\sigma}\rangle $ and
 $|c_{m\sigma}\rangle$  under the proviso that the f-state  $|f_{i\bar{\sigma}}\rangle$ is not occupied, whereas the second represents the part of the f-c quantum-mechanical mixing when the admixture of a doubly-occupied state with the opposite 
spins,  $|f_{i\sigma}\rangle$ and $ |f_{i\bar{\sigma}}\rangle$, is admissible. The various hopping processes involving the original (bare) atomic f electrons and the resulting occupation-resolved hybridization processes are depicted in Fig. \ref{fig:1} (for their detailed meaning see below).
Since the heavy fermion physics is related to the limit: $|\epsilon_f|\geq V$ and $|\epsilon_f|\ll \epsilon_f +U$ (with $\epsilon_f \equiv \epsilon_f-\mu$), the
first term in (\ref{eq:2}) corresponds to the low-energy mixing processes (represent a real f-c mixing,  Fig. \ref{fig:1}(b)), whereas the processes represented by the second term are realized only  via virtual (high-energy processes, Fig. \ref{fig:1}(c,d)) and are accounted for in the second-order in $V/(U+\epsilon_f)$ (the first nontrivial order). The latter
processes are removed from the original Hamiltonian (\ref{eq:1}) via canonical transformation proposed earlier \cite{Spalek1990151, PhysRevB.38.208}.
The original Hamiltonian is then transformed to the effective Hamiltonian of the form

\begin{eqnarray}
\mathcal{H} &=& \mathcal{\hat{P}} \left\{ 
\sum_{mn\sigma} \left(t_{mn} c^{\dagger}_{m\sigma} c_{n\sigma} -\sum_{i} \frac{V_{im}^{*} V_{in}}{U + \epsilon_f} \hat{\nu}_{if\bar{\sigma}}c^{\dagger}_{m\sigma} c_{n\sigma}  \right) 
 \right\} \mathcal{\hat{P}}  \nonumber\\
&+&
 \mathcal{\hat{P}} \left\{  \sum_{imn \sigma} \frac{ V_{im}^{*} V_{in} }{U + \epsilon_f}\hat{S}_{i}^{\sigma} c^{\dagger}_{m\bar{\sigma}} c_{n\sigma} +\sum_{i\sigma}{\epsilon_f } \hat{\nu}_{if\sigma}\right\}
 \mathcal{\hat{P}}\nonumber\\
&+&
 \mathcal{\hat{P}} \left\{ \sum_{im\sigma} (1-\hat{N}_{i\bar{\sigma}}) (V_{im} f^{\dagger}_{i\sigma} c_{m\sigma} +  H.c.) \right\}
 \mathcal{\hat{P}}\nonumber\\
 &+&
 \mathcal{\hat{P}} \left\{ 
\sum_{im\sigma} \frac{2 |V_{im}|^2
  }{U+\epsilon_f} \left(\mathbf{\hat{S}}_i\cdot \mathbf{\hat{s}}_m - \frac{\hat{\nu}_{if}\hat{n}_{m}}{4} + \frac{\hat{\nu}_{if\bar{\sigma}} \hat{n}_{m\sigma}}{4} \right)  \right\}
 \mathcal{\hat{P}}\nonumber\\
&+&
\mathcal{\hat{P}} \left\{ -\frac{1}{2} g_f\mu_B H \sum_{i\sigma} \sigma \hat{N}_{i\sigma} - \frac{1}{2} g_c\mu_B H \sum_{m\sigma} \sigma \hat{n}_{m\sigma}  \right\}
 \mathcal{\hat{P}}\nonumber\\ 
 &+&
 \mathcal{\hat{P}} \left\{ 
\sum_{im\sigma} J_{ij} \left(\mathbf{\hat{S}}_i\cdot \mathbf{\hat{S}}_j - \frac{\hat{\nu}_{if} \hat{\nu}_{jf}}{4} \right)  \right\}
 \mathcal{\hat{P}}
,
\label{eq:3}
\end{eqnarray}
where $t_{mn}\equiv t_{mn}-\mu$, $\epsilon_f \equiv \epsilon_f-\mu$. Also, in this Hamiltonian we have projected out completely the double occupancies of
f states, as well as have defined the following projected quantities for f electrons $\hat{\nu}_{if\sigma} \equiv \hat{N}_{i\sigma}(1-\hat{N}_{i\bar{\sigma}})$, $\hat{\nu}_{if}\equiv \sum_{\sigma}\hat{\nu}_{if\sigma}$, $\mathbf{\hat{S}}_i\equiv (\hat{S}_i^{\sigma}, \hat{S}^{z}_i )= [f^{\dagger}_{i\sigma} f_{i\bar{\sigma}}, 1/2 (\hat{\nu}_{if\uparrow}-\hat{\nu}_{if\downarrow})]$, and the corresponding (non-projected) quantities for the conducting band electrons are $\hat{n}_{m\sigma} = c^{\dagger}_{m\sigma}c_{m\sigma}$, $\hat{n}_m = \sum_{\sigma} \hat{n}_{m\sigma}$ and  $\mathbf{\hat{s}}_m\equiv (\hat{s}_m^{\sigma}, \hat{s}^{z}_i )= [c^{\dagger}_{m\sigma} c_{m\bar{\sigma}}, 1/2 (\hat{n}_{m\uparrow}-\hat{n}_{m\downarrow})]$. Additionally, as we assume that $g_c=g_f=g =2$, the applied field is defined as $h\equiv \frac{1}{2} g \mu_B H$. Finally, we have added an effective 4-th order term $\sim J_{ij}\sim \sum_{m} (V_{im}V_{jm})^4/(\epsilon_f+U)^3$, which will not be
written up explicitly here, as we concentrate only on the second-order effects. 
The Gutzwiller projector $\hat{\mathcal{P}} = \prod_{i} (1- \hat{N}_{i\uparrow}\hat{N}_{i\downarrow})$ eliminates explicitly the double 
f-occupancies (it can be dropped as the operators inside $\hat{\mathcal{P}}(...)\hat{\mathcal{P}}$ are already in the projected form). The first term in (\ref{eq:3}) contains c-c hoping (see Fig. \ref{fig:1}(a)) renormalized by the virtual-hopping process, the second term is effective hopping with the spin-flip exchange between c and f electrons (see Fig. \ref{fig:1}(b)), the third part and the fourth - the f-level energy and projected hybridization, while the fifth - the effective f-c antiferromagnetic Kondo coupling. The minimal 
model coming out of that procedure is to neglect an effective f-f exchange and the renormalization of the $c-c$ hopping. 
In result, we start from a simplified effective Hamiltonian which we rewrite in the explicit form          

 \begin{eqnarray}
\mathcal{H} &=& \mathcal{\hat{P}} \left\{ 
\sum_{mn\sigma} t_{mn} c^{\dagger}_{m\sigma} c_{n\sigma} + \sum_{im\sigma} \frac{2 |V_{im}|^2
  }{U+\epsilon_f} \frac{\hat{\nu}_{if\bar{\sigma}} \hat{n}_{m\sigma}}{4} 
 \right\} \mathcal{\hat{P}}  \\
&+&
 \mathcal{\hat{P}} \left\{\sum_{i\sigma}{\epsilon_f } \hat{\nu}_{if\sigma} + \sum_{im\sigma} (1-\hat{N}_{i\bar{\sigma}}) (V_{im} f^{\dagger}_{i\sigma} c_{m\sigma} +  H.c.) \right\}
 \mathcal{\hat{P}}\nonumber\\
 &+&
 \mathcal{\hat{P}} \left\{ 
\sum_{im\sigma} \frac{2 |V_{im}|^2
  }{U+\epsilon_f} \left(\mathbf{\hat{S}}_i\cdot \mathbf{\hat{s}}_m - \frac{\hat{\nu}_{if} \hat{n}_{m} }{4} \right)  \right\}
 \mathcal{\hat{P}}\nonumber\\
 &+&
 \mathcal{\hat{P}} \left\{ -\frac{1}{2} g_f\mu_B H \sum_{i\sigma} \sigma \hat{N}_{i\sigma} - \frac{1}{2} g_c\mu_B H \sum_{m\sigma} \sigma \hat{n}_{m\sigma}  \right\}
 \mathcal{\hat{P}}\nonumber.
\label{eq:4}
\end{eqnarray}
We see that we have simultaneously presented a residual (projected) hybridization term and the Kondo kinetic exchange interaction, with the exchange integral $J_K \equiv J_{im} = 2|V_{im}|^2/(U+\epsilon_f)$, which differs from that obtained with the help of the Schrieffer-Wolff transformation (as it contains only high-energy virtual-hopping processes, cf. Fig. \ref{fig:1}(b)). The appearance of a number of magnetic phases arises from a competition/cooperation of the projected hybridization and antiferromagnetic Kondo interaction.

\subsection{A general comment: Anderson, Kondo or mixed Anderson-Kondo lattice representation?}
\label{B}
It may be useful at this point to characterize briefly the subtle difference between the present formulation and the original Anderson- or Kondo-lattice models. First, our canonical transformation differs from the standard Schrieffer-Wolff transformation. Namely, in the Schrieffer-Wolff \cite{PhysRev.149.491} (and related Coqblin-Schrieffer \cite{PhysRev.185.847}) transformations the whole hybridization term is transformed out and replaced by the Kondo interaction. Here, only the second part of (\ref{eq:2}) represents virtual processes (cf.  Fig. \ref{fig:1}(b) ), whereas the first term of (\ref{eq:2}), cannot be transformed out, since $V \sim \epsilon_f$ and hence the corresponding processes lead to the itineracy of f-electrons (cf. Fig. \ref{fig:1}(d) for an illustration of an effective f-f hopping). This means that 
the f-electrons are not strictly localized, as in most of the HF systems they are not. As a consequence, the total number of f particles is not conserved, only the total number $\hat{N}_e = \sum_{i\sigma}\hat{N}_{i\sigma} +\sum_{m\sigma}\hat{n}_{m\sigma} $ of \textit{all} particles is a conserved quantity.
The last statement amounts to saying that the physics of Hamiltonian  (\ref{eq:3}) and (4) is contained in that coming of (\ref{eq:1}),
but in general, is not equivalent to that of \textit{true Kondo-lattice model}, i.e. when the third term in (4) $\sim V$ is absent.
This is similar to the difference between the physics coming from the Anderson impurity model as compared to that coming from the Kondo model; 
the later representing the asymptotic limit of the former \cite{PhysRevLett.35.1101}. 

In principle, one could say that all the relevant physics is fully contained in the general form (\ref{eq:1}). Why then introducing (\ref{eq:3}) or (4)
which represent its particular limiting forms? This is the first basic question.  The answer to this question is as follows. The form (\ref{eq:1}) represents indeed a general hybridized two-orbital system with the short-range (Hubbard) interaction (apart from the absence of Falicov-Kimball term $\sim \sum_{<im>}U_{im}\hat{N}_i \hat{n}_m$,  the role of the missing Falicov-Kimball term is  discussed in Appendix \ref{A2}). No exact solution is available in the lattice situation of dimensionality $D>>1$. Therefore, we have to resort to approximations and it is usually profitable, and even physically plausible, to take into account the principal interorbital/intersite exchange interaction explicitly first and carry out
a simplified (e.g. mean-field like) analysis on such an effective Hamiltonian subsequently, with the most relevant correlations included, even in the approximate manner in the mean-field type approximation. What is equally important,
the form of the effective Hamiltonian, by containing physical (low-energy) interactions, helps in selecting relevant order parameters and related to them mean 
fields in nontrivial cases, as discussed in detail below in one such particular situation. Parenthetically, the analysis is of the same type as that when transforming the Hubbard model into an effective $t-J$ model and determining nontrivial many body ground states for the latter. Such Hamiltonian is applicable to the analysis of both normal and superconducting states, the latter with real space pairing included.

One has to note that for the Kondo-lattice Hamiltonian, which has general form 

\begin{equation}
\mathcal{H}_{KL} = \sum_{mn\sigma} t_{mn} c^{\dagger}_{m\sigma} c_{n\sigma} +\sum_{im} J_{im} \mathbf{\hat{S}_i} \cdot \mathbf{ \hat{s}_m }
+\sum_{im} J^{'}_{ij} \mathbf{ \hat{S}_i }\cdot \mathbf{\hat{S}_j },
\label{eq:5}
\end{equation}    
in which the numbers of f-particles and c-particles are conserved \textit{separately}, i.e. both $\hat{N}_f \equiv \sum_{i\sigma} \hat{N}_{i\sigma}$ and 
$\hat{N}_c \equiv \sum_{m\sigma} \hat{n}_{m\sigma}$ commute with $\mathcal{H}_{KL}$ separately. Therefore, in discussing the HF or superconducting states
based on this type of Hamiltonian \cite{1742-6596-200-1-012162}  one has to assume that those two global conservation laws are only approximately obeyed. It is better to use our effective Hamiltonian (\ref{eq:3}) or (4), on the expense of the simplicity though.

A second basic question still remains and concerns the reduction of Hamiltonian (4) or (\ref{eq:3}) to (\ref{eq:5}) in a consistent manner. This is usually done by removing the residual hybridization term and completing the Schrieffer-Wolff transformation. However, in our mean-field
analysis the true Kondo-lattice limit expressed by (\ref{eq:5}), is effectively achieved as a limit $n_f\rightarrow 1$. This limit is indeed achieved when $V/\epsilon\rightarrow 0$, as we show explicitly in the next Section in concrete situation.

\subsection{Efective mean-field Hamiltonian  and statistically-consistent mean-field approximation}
  
In order to construct a mean field Hamiltonian we assume, that a proper ground variation state
can be represented in the form $|\Psi\rangle = \mathcal{\hat{P}}|\Psi_0\rangle$, where 
$|\Psi\rangle$  indicates the correlated state and $|\Psi_0\rangle$ is an eigenstate of the effective single-particle Hamiltonian.
Then the expectation value of any operator can be  in principle calculated as

\begin{equation}
\langle \hat{\mathcal{O}}\rangle \equiv \frac{\langle \Psi|\hat{\mathcal{O}}|\Psi\rangle}{\langle\Psi|\Psi\rangle} =  \frac{\langle \Psi_0|\hat{\mathcal{P}}\hat{\mathcal{O}}\hat{\mathcal{P}}|\Psi_0\rangle}{\langle\Psi_0|\hat{\mathcal{P}}^2|\Psi_0\rangle} \equiv \frac{\langle \hat{\mathcal{P}}\hat{\mathcal{O}} \hat{\mathcal{P}}\rangle_0}{\langle\hat{\mathcal{P}}^2\rangle_0}.
\label{eq:6}
\end{equation}
In most cases, the above calculations  represent  a non trivial task, mostly because the projection operator is non-local and by applying the Wick theorem 
many terms appear in the process of Hartree-Fock type decoupling. To carry out these evaluations  an approximation is necessary at this point. Here, we use the scheme proposed
recently by Fukushima \cite{PhysRevB.78.115105} in the local-constraint version, which assumes
that the average number of correlated f-electrons at each site and with an arbitrary spin orientation is unchanged
by the projection, i.e. $\langle \hat{N}_{i\sigma}\rangle = \langle \hat{N}_{i\sigma}\rangle_0 \equiv n_{if\sigma}$.
In the subsequent analysis, we use more general Gutzwiller-type projector, $\mathcal{\hat{P}} = \prod_{i} \lambda_{i\uparrow}^{\hat{N}_{i\uparrow}/2} 
\lambda_{i\downarrow}^{\hat{N}_{i\downarrow}/2}(1-\hat{N}_{i\downarrow}\hat{N}_{i\uparrow} )$, which eliminates the double occupancies of f-electrons in real space and conserves average number of f-electrons at each site before and after projection; the latter is accomplished by fixing the fugacity factors $\{\lambda_{i\sigma}\}$. 
In order to derive the explicit form of the fugacity factors, one  has to calculate the density of $\sigma$ spin f-electron at site i, namely, $\langle \hat{N}_{i\sigma}\rangle = \frac{\langle \lambda_{i\sigma} \hat{N}_{i\sigma} (1-\hat{N}_{i\bar{\sigma}}) \prod_{l \neq i} \mathcal{\hat{P}}^2_l \rangle_0} {\langle\prod_{l} \mathcal{\hat{P}}^2_l \rangle_0}$. This can be done  by
taking only the onsite contraction, as the intersite ones are much smaller. Then, $l\neq i$ terms cancel out (in the
numerator with these of the denominator). Therefore, the condition to determine the fugacity factors is given by $\langle \hat{N}_{i\sigma}\rangle = \frac{ \lambda_{i\sigma} (1 -  \langle {N}_{i\bar{\sigma}}\rangle_0 )  \langle {N}_{i\sigma}\rangle_0 }{\langle \mathcal{P}^2_i\rangle}$, which results in the explicit expression for the fugacity factors,  $\lambda_{i\sigma} = \frac{1-n_{if\sigma}}{1-n_{if}}$\cite{PhysRevB.78.115105}.

It should be noted that
the uncorrelated c-electrons are not effected by the projection procedure, as it projects out only for the number of f-electrons.
 Though the original local-constraint scheme for calculating the correlated averages was proposed for a single band Hubbard model \cite{PhysRevB.78.115105} and was successfully  applied in $t-J$ model \cite{PhysRevB.83.104512}, it can be easily extended to the periodic Anderson model in a straightforward manner (see Appendix \ref{A0} for details).

In effect, this kind of Gutzwiller approximation yields  the mean value of the Hamiltonian (4) 
\begin{eqnarray}
\frac{\langle\mathcal{H}\rangle}{\Lambda} &=&  8 t \xi -  \frac{V^2}{U +\epsilon_f}
\sum_{\sigma=\uparrow,\downarrow} n_{f\bar{\sigma}} n_{c\sigma} \nonumber \\
  &-&  \frac{V^2}{U +\epsilon_f}
\sum_{\sigma=\uparrow,\downarrow}\left(\frac{n_{f\bar{\sigma}}}{1-n_{f\sigma}} \gamma^*_{\sigma} \gamma_{\sigma}
\right)  \nonumber \\
&+&  \sum_{\sigma=\uparrow,\downarrow} V \sqrt{\frac{1-n_f}{1-n_{f\sigma}}} \left(\gamma_{\sigma} + \gamma^*_{\sigma}\right) \nonumber\\
&-&\frac{V^2}{U + \epsilon_f} \sum_{\sigma=\uparrow,\downarrow}\frac{ \gamma_{\sigma}\gamma^*_{\bar{\sigma}}}{\sqrt{(1-n_{f\sigma})(1- n_{f\bar{\sigma}})}} \nonumber \\
&+& \epsilon_f n_f - h_f m_f - h_c m_c,  
\label{eq6}
\end{eqnarray}
where $\Lambda$ is the number of lattice sites and we have additionally assumed, that all the mean fields appearing in our model represent spatially homogeneous quantities, namely, i. e.,  
\begin{eqnarray}
&&n_{if} = \sum_{\sigma=\uparrow,\downarrow } \langle \hat{N}_{i\sigma}\rangle \equiv  n_f,\ n_{mc} = \sum_{\sigma=\uparrow,\downarrow} \langle \hat{n}_{m \sigma}\rangle \equiv  n_c\nonumber,\\
&&m_{if} = \frac{1}{2} \sum_{\sigma=\uparrow,\downarrow} \langle \sigma \hat{N}_{i\sigma}\rangle \equiv m_f, \nonumber\\
&& m_{mc} = \frac{1}{2}\sum_{\sigma=\uparrow,\downarrow} \langle \sigma \hat{n}_{m\sigma}\rangle \equiv m_c,\nonumber\\
&& \gamma_{im\sigma} = \langle f^{\dagger}_{i\sigma}c_{m\sigma}\rangle_0 \equiv \gamma_{\sigma}, \ 
\xi_{mn\sigma} = \langle c^{\dagger}_{m\sigma}c_{n\sigma}\rangle_0 \equiv \xi. 
\end{eqnarray}
 
It should be noted that the hybridization term in (\ref{eq6}) has the same renormalization  factor as can be obtain in standard Gutzwiller approximation \cite{PhysRevLett.55.995,PhysRevB.34.6420} in the $U=\infty$ limit.

Also, in order to formulate a completely correct mean-field approximation, we supply the mean-field Hamiltonian (\ref{eq6})
with constrains via the Lagrange-multiplier method. By doing so we ensure the statistical consistency of the present variational method \cite{PhysRevB.81.073108,PhysRevB.83.104512} of evaluating the averages in the sense, that they also coincide with those calculated via a fully self-consistent procedure. This method provides results which are equivalent to those obtained in the saddle-point approximation within the slave-boson approach \cite{PhysRevB.58.3293}. In result, the effective MF Hamiltonian takes the form

\begin{eqnarray}
\mathcal{H}_{MF} &=& -\sum_{<mn>\sigma} \left(\eta (c^{\dagger}_{m\sigma}c_{n\sigma}
 - \xi) + H.c.\right) \nonumber\\
 &-& \lambda_{cm} \sum_{m\sigma}\left( \sigma c^{\dagger}_{m\sigma}c_{m\sigma}-m_c \right)\nonumber\\
 &-&  \mu \sum_{i\sigma}\left( c^{\dagger}_{i\sigma}c_{i\sigma}+f^{\dagger}_{i\sigma}f_{i\sigma} \right)\nonumber\\ 
&-&   \sum_{i\sigma}\left( \tau_{\sigma}(f^{\dagger}_{i\sigma}c_{i\sigma} - \gamma_{\sigma})+ H.c.\right) \nonumber\\ 
&-&  \lambda_{f} \sum_{i\sigma}\left(  f^{\dagger}_{i\sigma}f_{i\sigma}-n_f \right) - \lambda_{fm} \sum_{i\sigma}\left(\sigma  f^{\dagger}_{i\sigma}f_{i\sigma}-m_f \right)\nonumber\\
&-& \lambda \sum_{i\sigma}\left(  f^{\dagger}_{i\sigma}f_{i\sigma} + c^{\dagger}_{i\sigma}c_{i\sigma}-n_e \right) \nonumber\\
&+& \langle\mathcal{H}\rangle,
\label{eq8} 
\end{eqnarray}   
where $<m,n>$ means the summation over nearest neighboring pairs. Note that the field $\lambda_{fm}$ may play role of a molecular field induced by electronic correlations whereas $\lambda_f$ and $\lambda$ represent shifts of the chemical potential (the role of $\lambda$ proves to be minor). 

Now, Eq. (\ref{eq8}) has a particularly simple $\mathbf{k}$-space representation:

\begin{equation}
\mathcal{H}_{MF} = \sum_{\textbf{k}\sigma} 
\begin{pmatrix} 
c^{\dagger}_{\textbf{k} \sigma} &  f^{\dagger}_{\textbf{k} \sigma}  \\
\end{pmatrix} 
\begin{pmatrix}
\epsilon^c_{\textbf{k} \sigma} & -\tau_{\sigma}\\
-\tau_{\sigma} & \epsilon^f_{\sigma}\\
\end{pmatrix}
\begin{pmatrix} 
c_{\textbf{k} \sigma} \\
f_{\textbf{k} \sigma} \\
\end{pmatrix}
+ C,
\label{eq9}
\end{equation}
where $n_e = n_f+n_c$ is the total number of electrons per site and the constant  $C$ includes all the wavevector independent quantities: $C \equiv \langle\mathcal{H}\rangle + \Lambda(16 \xi\eta
+ \lambda_{cm} m_c + 2\sum_{\sigma}\tau_{\sigma}\gamma_{\sigma} + \lambda_f n_f + \lambda_{fm} m_f + \lambda n)$. For example, the band energy of c-electrons in a two-dimensional case is represented explicitly by $\epsilon^c_{\textbf{k}\sigma} \equiv -4\eta \left(\cos{ k_x} + \cos{k_y}\right) - \mu - \lambda -\lambda_{cm} \sigma$ and the energy of the localized f-electrons equals $\epsilon^f_{\sigma} \equiv -\mu- \lambda -\lambda_f - \lambda_{fm}\sigma $.
It is easy to see that the eigenvalues of the Hamiltonian (\ref{eq9}) are then
\begin{equation}
E_{k\sigma}^\alpha = \frac{1}{2}\left(\epsilon_{k\sigma}^c + \epsilon^f_{\sigma} + \alpha \sqrt{(\epsilon_{k\sigma}^c - \epsilon^f_{\sigma})^2 + 4\tau_{\sigma}^2}\right), 
\label{eq10}
\end{equation}
where $\alpha=\pm 1$ labels the quasiparticle band (antibonding and bonding components, respectively). The quasiparticles form two bands with renormalized characteristics.

The mean-field values can be determined by minimizing the free energy potential $\mathcal{F} \equiv \mathcal{F}_0 - \frac{1}{\Lambda\beta} \sum_{\textbf{k}\sigma \alpha } \ln{\left(1+e^{-\beta E_{\textbf{k}\sigma}^\alpha }\right)}$, i.e. by solving 
the following system of equation representing the necessary conditions for the minimum:

\begin{eqnarray}
\frac{\partial\mathcal{F}}{\partial n_e} = 0,  &\qquad & \frac{\partial\mathcal{F}}{\partial n_f} = 0, \qquad   \frac{\partial\mathcal{F}}{\partial m_c} = 0\nonumber,\\
 \frac{\partial\mathcal{F}}{\partial m_f} = 0,  &\qquad& \frac{\partial\mathcal{F}}{\partial \gamma_{\sigma}} = 0, \qquad  
\frac{\partial\mathcal{F}}{\partial \xi} = 0\nonumber,\\
 \frac{\partial\mathcal{F}}{\partial\lambda_i} = 0,
\end{eqnarray}
where  equations with $\lambda_i=\{ \lambda,\lambda_f,\lambda_{fm}, \lambda_{cm}, \eta, \tau_{\sigma}\}$ can be easily solved analytically,
so that we are left with six equations which we solve numerically with the help of GSL (Gnu Scientific Library),
on the lattice of  $\Lambda = 256\times 256 $ sites, using the periodic boundary conditions, with the following values of the  parameters: $\epsilon_f = -0.75 W$, $t =1/8 W$, $U =5 W$, and for sufficiently low temperature $k_B T/ W= 5\cdot 10^{-5}$, where $W$ is the  bare conduction-electron bandwidth, that sets our energy scale ($W = 1$). Exemplary values obtained via the self-consistent and variational method for a ferromagnetic state and $h=0$ are displayed in Table \ref{tab:1}. The numerical accuracy is at least on the last digit of the quantities displayed in that Table. Also the numerical results are illustrated further on the example of a two-dimensional (square) lattice. Parenthetically, a detailed shape of the density of states in bare band is of secondary importance.

\begin{table}
\caption{\label{tab:1} Values of the chemical potential and the MF parameters. The physical free energy $F = \mathcal{F} + \mu n_e$ for $n_e = 1.9 $ and $1.75 $ is  $-1.04971$ and $-1.04175$, respectively. We see that in this limit the f-electron occupation is close to unity and their magnetic moment ($m_f$) is decisively larger than that of the compensating Kondo cloud ($m_c$). }
\begin{ruledtabular}
\begin{tabular}{|l|l|l|l|l|l|l|}
\multicolumn{7}{|c|}{$U = 5$, $e_f = -0.75$, $|V| = 0.3$ } \\
\hline
  quantity & $\gamma_{\uparrow}$& $\gamma_{\downarrow}$ & $\mu$ & $n_f$ & $m_c$ & $m_f$ \\ \cline{1-7}
$n_e= 1.75$  & 0.4287 & 0.3182 & -0.0781 & 0.8621 & -0.0479 & 0.1768   \\ \cline{1-7}
$n_e= 1.9$  & 0.4278 & 0.3798 & -0.0291 & 0.8691 & -0.0245 & 0.0784    \\ \cline{1-7}
  quantity & $\tau_{\uparrow}$ & $\tau_{\downarrow}$ & $\lambda$ & $\lambda_f$ & $\lambda_{fm}$&  $\lambda_{cm}$   \\ \cline{1-7}
$n_e= 1.75$    & 0.1962 & 0.1512  & 0.0091& 0.0451 & 0.1450 & -0.0074 \\\cline{1-7}
$n_e= 1.9$    & 0.1765 & 0.1578  & 0.0092 & -0.0188 & 0.0598 & -0.0032
\end{tabular}
\end{ruledtabular}
\end{table}

\section{Results and Discussion}
\label{results} 

\subsection{Phase diagram and principal characteristics of selected phases}

\begin{figure}
 \includegraphics[width=0.5\textwidth]{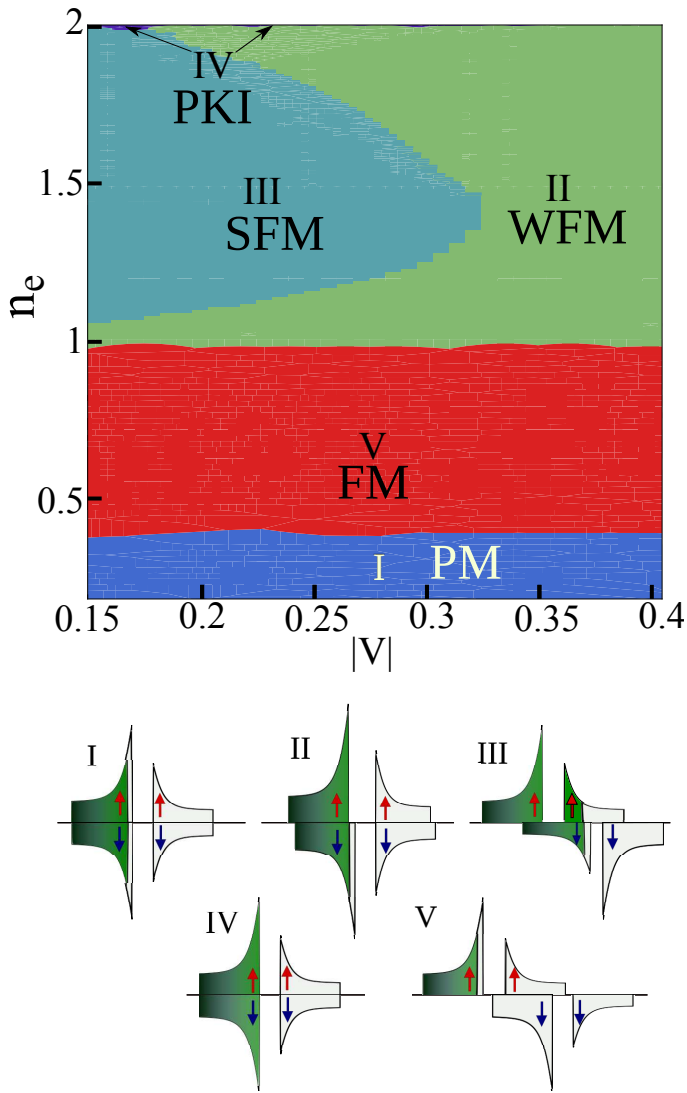}                
  \caption{ (Color online) Phase diagram (top) on $n_e-|V|$ plane displaying various arrangements of quasiparticle spin subbands (bottom) relative to the chemical potential.  In phase II  (weakly ferromagnetic metallic - WFM) the lower majority-spin subband is fully occupied, whereas the minority-spin subband remains partially filled.  
 The onset of region II is marked by the disappearance of the spin-up Fermi surface sheet, as the chemical potential is placed
in the hybridization gap and the lower spin-up subband is completely filled. For the small values of hybridization, the majority-spin electrons fill the upper hybridized band, what results in a full polarization of the system (phase III - strongly feromagnetic metallic SMF). The case (IV) represents the paramagnetic Kondo insulator (PKI) and can be observed only for $n_e = 2.0$. The phase diagram is drawn for $U = 5$, $\epsilon_f = -0.75$, and $t = -1/8$, all in units of the conduction electron bandwidth $W$. The difference between various ferromagnetic phases is connected with the difference in the shape and the filling of various parts of the density-of-states, as shown at the bottom.}
 \label{fig:phases}
\end{figure}

\begin{figure}
  \centering
 \includegraphics[width=0.5\textwidth]{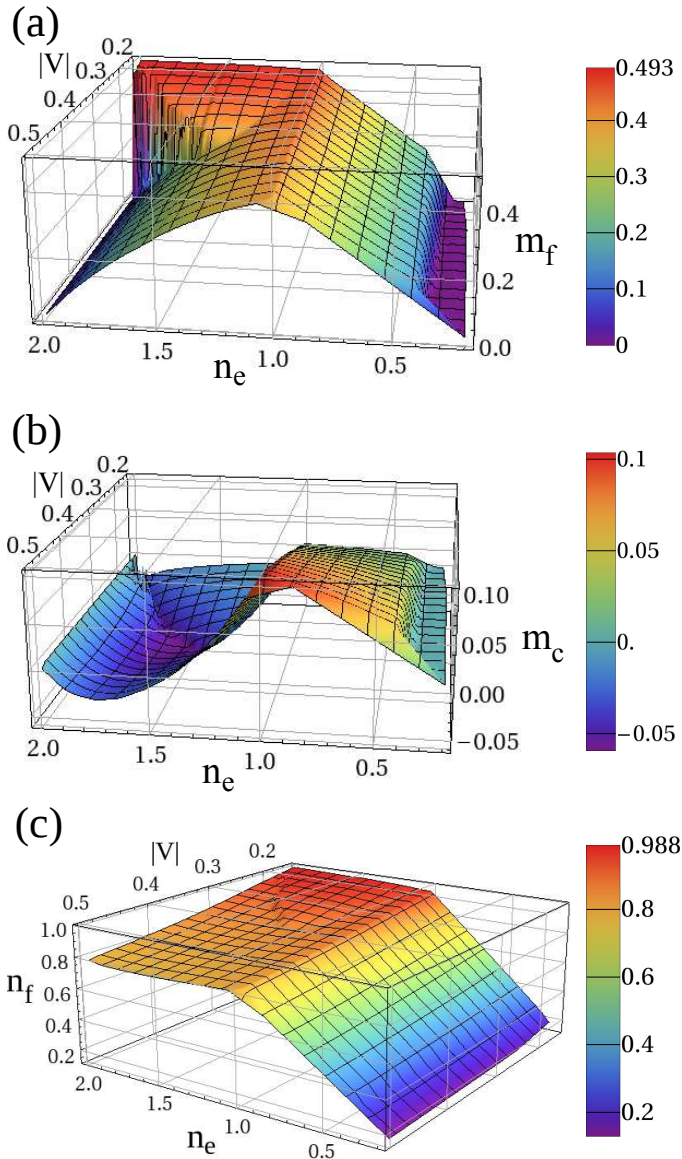}                
  \caption{(Color online) The magnetic moments of f-electrons (a) and of conduction electrons (b) as function of bare hybridization magnitude $|V|$ and total number of electrons per site: $n_e = n_c + n_f$, and (c) the number f-electron occupancy. Other  parameters are the same as those listed previously. }
\label{fig:mfmc}
\end{figure}

    A  variety of magnetic phases appears in this model depending on the value of microscopic parameters and the type of long-range order obtained. We analyze in detail only the phases specified in Fig. \ref{fig:phases} (bottom part) as well as provide the phase diagram on the $n_e-V$ plane (top part). For the case when the hybridization exists only between the electrons and holes at the same site, i.e. $V_{im} = V \delta_{im}$ and is $\mathbf{k}$-independent, we can precisely evaluate the effective mass of quasiparticles using analytical results (see Appendix \ref{A1} for details). It can be seen from Fig. \ref{fig:phases}  that, depending on the number of total electrons $n_e$ and the value of hybridization parameter $|V|$, we can achieve different states. In phase II,  which corresponds to the weakly ferromagnetic metal (WFM), the lower majority-spin subband is fully occupied, whereas the minority-spin subband remains partially filled.  
The onset of region II is marked by the disappearance of the spin-up Fermi-surface sheet, since the chemical potential is placed then in the hybridization gap and the lower spin-up subband is
completely filled. If the lower majority-spin subband in phase II is fully occupied (contains one electron per site)
then the minority-spin subband remains partially filled and contains $n_e-1$ electrons per site, what results in total
magnetization $ m = (2-n_e)/2$ and approaches zero for $n_e \rightarrow 2$ (cf. Fig. \ref{fig:mfmc}(a,c) and phase II in the Fig. \ref{fig:phases}). On the other hand, for a small value of hybridization the majority-spin electrons in the phase III  start filling the higher hybridized band, what results in  an almost full polarization of the system (see Fig. \ref{fig:mfmc}(a,c)). As the magnetizations of f- and c-electrons, $m_f$ and $m_c$, have  opposite signs due to antiferromagnetic  Kondo coupling between them, the maximum value of the total magnetization never exceeds $0.5\mu_B$ i.e. the maximum value in the localized limit.

\begin{figure}[h]
  \centering 
 \includegraphics[width=0.5\textwidth]{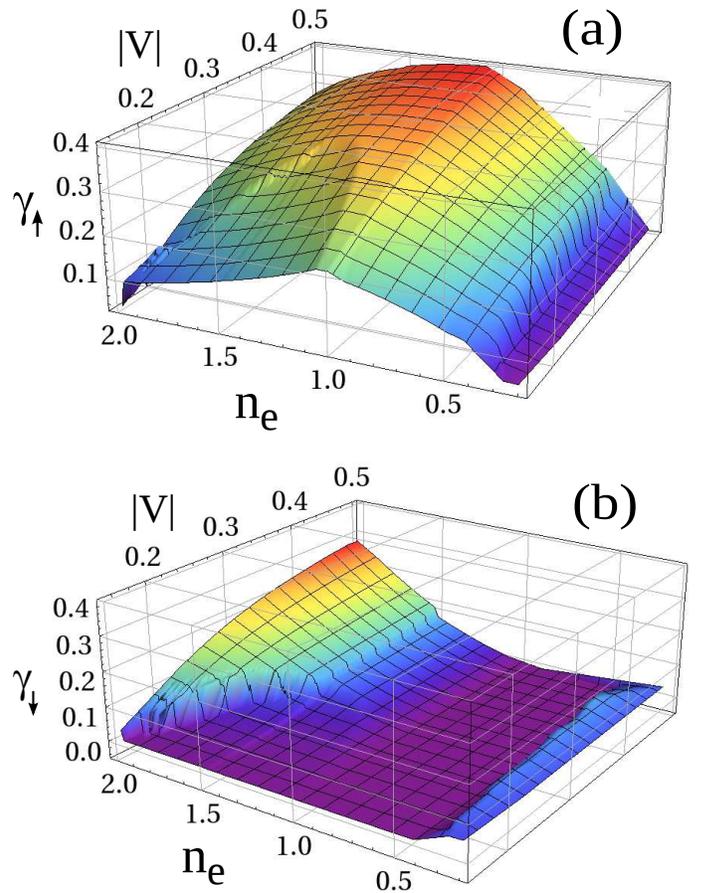}                
 \caption{(Color online) The mean magnitude of hybridization correlation functions: $\gamma_{\uparrow} = \langle f^{\dagger}_{i\uparrow}c_{i\uparrow}\rangle$ (a) and $\gamma_{\downarrow} = \langle f^{\dagger}_{i\downarrow}c_{i\downarrow}\rangle$ (b) both, as a function of total number of electrons and bare hybridization parameter $|V|$. Other parameters are the same as those used previously.}
  \label{fig:g}
\end{figure} 

In the case of $n_e<1.0$ both f- and c-electrons have the same sign of magnetization, as a consequence of the fact that the effective hybridization, represented by the mean-field value of $\gamma_{\downarrow} = \langle f^{\dagger}_{i\downarrow}c_{i\downarrow}\rangle$, is close to zero, as can be seen in Fig \ref{fig:g} b.  In the case when $n_e>1.0$ and the bare hybridization $|V|<0.3$, the f-electrons start localizing, as can be seen by a rapid increase of $m_f$ (see Fig. \ref{fig:mfmc}(a)) and the suppression of the hybridization between localized and itinerant electrons (in Fig. \ref{fig:g} see a recess of both $\gamma_{\downarrow}$ and $\gamma_{\uparrow}$ in the region corresponding to the phase III (SFM)). The Kondo singlet correlation are destroyed to yield a strong ferromagnetism. Thus large magnetic moments reflect the tendency to destroy the partial Kondo effect. Similar conclusions can be drawn from the $S=1$ underscreened Anderson lattice model \cite{PhysRevB.83.014415, PhysRevB.76.125101}. Here, the same effect is observed in orbitally nongenerate PAM.

\subsection{Ferromagnetic phase: suppression of heavy quasiparticle masses}
\begin{figure}
 \includegraphics[width=0.5\textwidth]{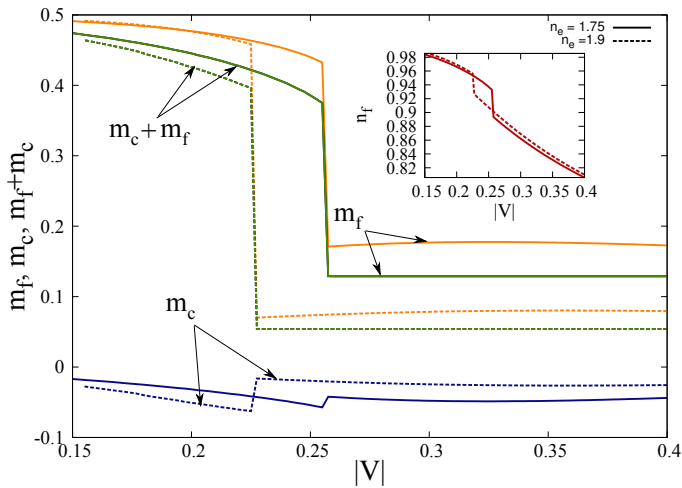}                
  \caption{(Color online) The magnetization of c- and f-electrons, $m_c$ and $m_f$, respectively, plotted as a function of hybridization parameter for selected numbers electron occupancy $n_e$. The inset shows the corresponding dependence of the occupancy $n_f$.  }
  \label{fig:mV}
\end{figure}

The detailed phase diagram should encompass also antiferromagnetic phases \cite{PhysRevB.58.3293}. In the present paper we shall limit our discussion to the range of parameters, for which the ferromagnetic phases II and III  are stable. The phase transition from the state with relatively small magnetic moments (phase II) to the phase with almost fully polarized spin-up subband (III) can be induced by  reducing the hybridization strength $|V|$, as can be seen in Fig. \ref{fig:mV}. We find, in agreement with DMFT \cite{PhysRevB.64.052402, PhysRevB.62.5657} and slave boson \cite{PhysRevB.58.3293} calculations, that there is 
a critical value of hybridization strength (cf. Fig. \ref{fig:mV}) below which the system becomes fully polarized (phase III). 
For large value of hybridization, the strong ferromagnetic phase is destroyed because of the screening the f-electron magnetic moments partly by the c-electrons and partly by the itineracy of f-electrons. While a strong hybridization increases the magnetic interaction, it simultaneously weakens the magnitude of magnetic moments by the itineracy. When the hybridization parameter $|V|$ falls below the critical value system transforms in a discontinuous manner from the strong ferromagnetic phase III to the weak ferromagnetic phase. In the inset to Fig. \ref{fig:mV} one can observe that the occupancy $n_f$  simultaneously undergoes 
a sharp transition as a function of the hybridization strength $|V|$. The limit $m_f\rightarrow 1/2$ and $n_f\rightarrow 1$ (for $|V|<0.15$) 
can be thus called the localized moment limit, in which indeed $V/\epsilon_f\ll1$.  

\begin{figure}
 \includegraphics[width=0.5\textwidth]{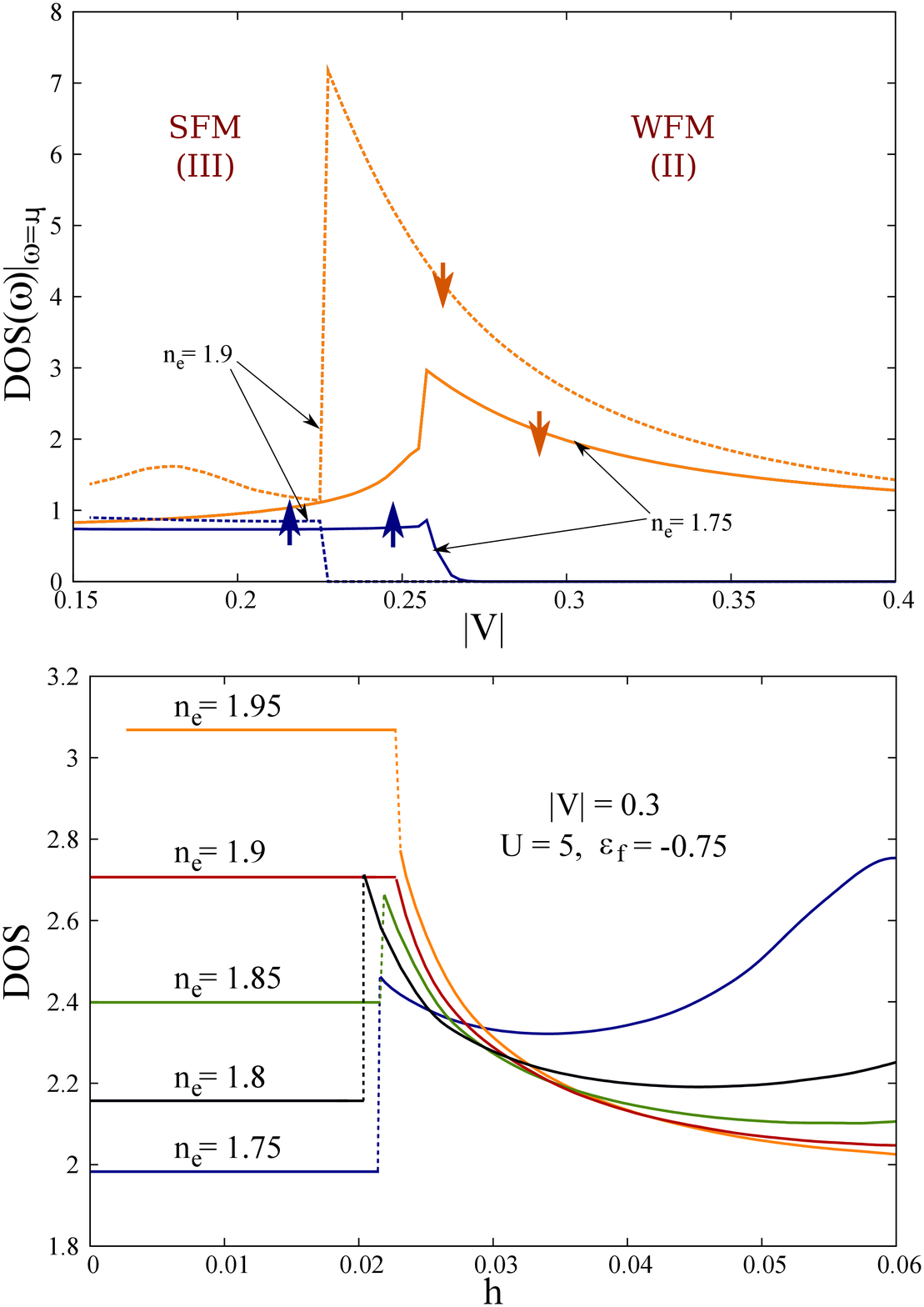}                
  \caption{(Color online) Density-of-states DOS enchancement at the Fermi level as a function of hybridization (a) and applied field (b), illustrating the nature phase transition from phase II to III. The parameters are
the same as those used above.  The arrows indicate the spin orientations.}
  \label{fig:dos}
\end{figure}

\begin{figure}
  \centering
 \includegraphics[width=0.5\textwidth]{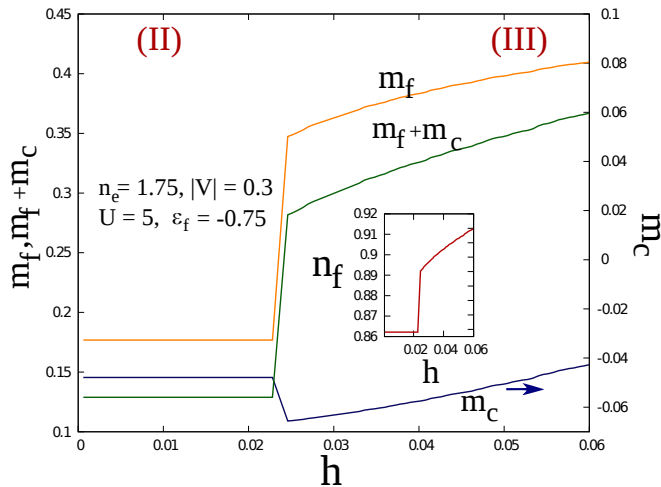}                
  \caption{(Color online) The magnetizations of $m_c$ and $m_f$, plotted as a function of applied magnetic field. The metamagnetic (discontinuous) transition reflects the critical changeover when crosing the regime WFM$\rightarrow$SFM. Inset: f-band occupancy change accompanying the metamagnetic transition.}
  \label{fig:mb}
\end{figure}


\begin{figure}
  \centering
 \includegraphics[width=0.5\textwidth]{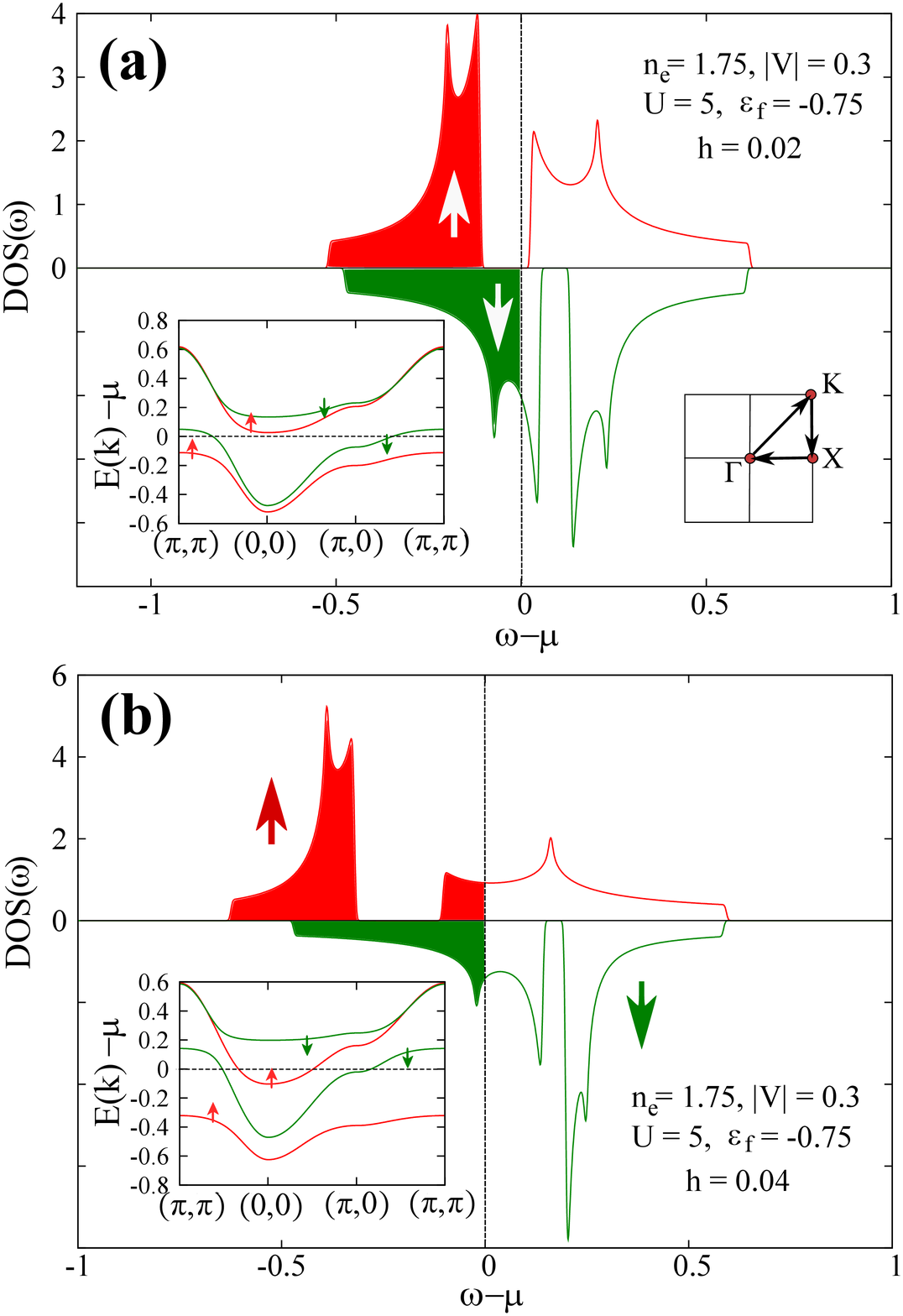}                
  \caption{(Color online) Spin resolved quasiparticle density of states (DOS) in WFM (a) and SFM (b) phases. The upper(lower) plots corresponds respectively to the spin up(down) electrons. Insets: the band substructure for specific parameters. The arrows label the spin orientation of the quasiparticles. The system is metallic in both WFM and SFM phases, but there is a change of spin polarization at Fermi energy. }
  \label{fig:d}
\end{figure}

The density-of-states (DOS) enhancement at the Fermi level is a direct measure of the effective mass enhancement of quasiparticles \cite{springerlink:10.1140/epjb/e2009-00102-y, PhysRevB.73.045117, PhysRevB.76.035119}.
As can be seen from Fig. \ref{fig:dos}, in the case of stable state II the density of states for the  spin-down subband increases with the increasing the hybridization and that the spin-up part is absent, because the chemical potential is placed in hybridization gap for the spin-up subbands. 

The phenomenon such as the metamagnetic behavior of the magnetization curve, as well as the field- and spin-dependent 
effective masses have been observed in several heavy fermion compounds \cite{PhysRevLett.71.2110, PhysRevLett.82.3669, McCollam20051}. For selected parameters, our model predicts that the  metamagnetism appears in the crossover region in between the regimes II to III (see Fig. \ref{fig:mb}). Namely, the principal feature of the heavy fermion state II (WFM phase) is its insensitivity to the strong applied magnetic field \cite{PhysRevB.77.094419}. The only quantity which changes significantly is the chemical potential position. With increasing value of magnetic field, the chemical potential shifts and eventually leaves the quasiparticle-gap regime (cf. Fig. \ref{fig:d}), entering into the spin-up upper hybridized band and switching the system to phase III.
This phase transition causes sudden increase of the magnetization and rearrangement of the quasiparticle subbands as can be seen in the insets to Fig. \ref{fig:d}. In the inset to Fig. \ref{fig:mb} one can see that define jump of $n_f$ in the applied magnetic field is observed while system evolves from weakly polarized (II) towards strongly polarized phase (III). The present result is consistent with the existing considerations, where additionally the Coulomb interaction between local and conducting electrons were taken into account  \cite{JPSJ.79.124712, PhysRevLett.100.236401}. The inclusion of the inter-orbital Coulomb interaction amplifies the sharpness of the transition.

\begin{figure}
  \centering
 \includegraphics[width=0.5\textwidth]{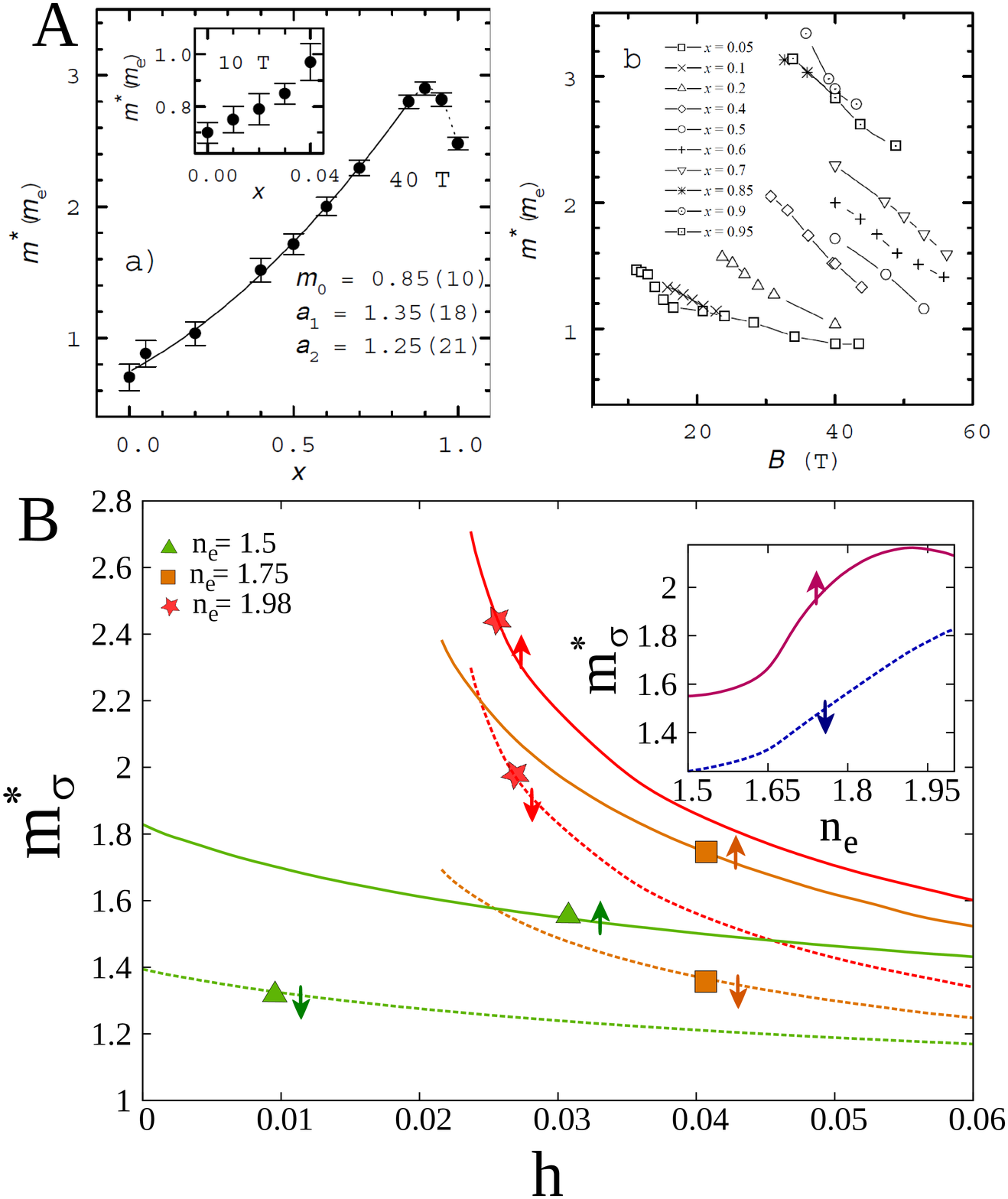}                
  \caption{(Color online) A: A plot of the quasiparticle effective mass of the
$\alpha_3$ frequency versus concentration at 40 T (left of top figure A), the field dependence of the effective mass 
for different concentrations (right of top figure A). Figure A represents experimental results for Ce$_x$La$_{1-x}$B$_6$ \cite{PhysRevLett.82.3669, Harrison2001234}. Note that for A the parameters specified the mass enhancement is rather small. B: The field dependence of the quasiparticle effective masses for $|V|=0.3$, remaining parameters are the same, as used before. The inset shows the quasiparticle effective masses (in units of bare mass) and their dependence from the total number of electrons $n_e$; for applied magnetic field $h = 0.03$ and $|V| = 0.3$. }
  \label{fig:mass}
\end{figure}

To understand better  the behavior of strongly polarized HF system in phase III we analyze also the field dependence of 
the quasiparticle mass. As can be seen in the Fig. \ref{fig:mass}, the quasiparticle mass is suppressed with increasing magnetic field. 
It can be interpreted as a destruction of the HF state. The magnetic field tends to order of both electrons and in this manner to suppress the  antiferromagnetic Kondo coupling. The effective mass reduction at high fields is  one of the characteristic features of HF \cite{JPSJ.73.769}. In contrast to the paramagnetic ground state of PAM \cite{JPSJ.77.023703}, we observe a completely different  behavior of the effective-mass field dependence. The first difference is that spin-up mass is larger than the spin-down carrier mass due to the discussed above rearrangements of quasiparticle the subbands (see Fig. \ref{fig:d}), which is also consistent with the overall mean-field theory of Wasserman et. al. \cite{0953-8984-1-16-002} (who does not have the spin-split masses though).  Secondly, both effective spin-dependent masses ($\sigma = \pm 1$) decrease in the stronger applied field.  The effective mass of quasiparticles depends also from the total number of electrons $n_e$. In the  case of the fixed hybridization strength  the value of localized electrons $n_f$ is almost insensitive to the change in the total number of electrons, since in the region with an almost localized f-electrons ($n_f\rightarrow 1$), changes in $n_f$ are minor. Thus the $n_e$ dependence of effective quasiparticle mass in that region is mainly due to change in number of conduction electrons $n_c$. 

\subsection{Application to Ce$_x$La$_{1-x}$B$_6$}
The obtained  behavior can account for the field dependence of the quasiparticle effective mass of Ce$_x$La$_{1-x}$B$_6$ \cite{PhysRevLett.82.3669, Harrison2001234} in very strong fields. In this material, the dHvA signal is found to originate only from a single-spin-orientation Fermi surface  sheet. However, it was difficult to  verify, from which spin channel it comes from. From the experimental observation \cite{Harrison2001234} it seems that spin up channel progressively dominates the dHvA signal as the concentration $x$ is increased (i.e. the number $n_e$ increases). The experimental investigation claim that spin-up component may be dominant in this compound. In our approach we reproduce two important experimental features: (i) the  field suppression of effective masses at high magnetic field, (ii)  the non-monotonic $x$ dependence of $m^*$ for the dominant spin component (cf. inset to Fig. \ref{fig:mass}) and its small value for this heavy-fermion compound.

One has to mention that Ce$_x$La$_{1-x}$B$_6$ shows the Fermi $T^2$ - liquid behavior in low-temperature resistivity  \cite{PhysRevLett.97.237204}
even though it exhibits a complicated phase diagram on the magnetic field-temperature plane \cite{PhysRevLett.97.237204,Tayama199932,PhysRevB.55.8339}. The discussed here effective mass
suppression was observed in very strong fields, when the complicated quadrupolar  phases discussed in \cite{PhysRevLett.97.237204} have already disappeared. This is the reason why we can apply our result to this complicated system in strong fields.

\section{Summary and outlook}
\label{summary}
In the paper we have analyzed the effect of spin-split and applied-field dependent quasiparticle masses (or quasiparticle density of states) and their influence on selected magnetic properties. The results differ from those 
obtained for a single narrow band model. What is equally important, the heavy masses are strongly suppressed at the border of weakly- to strongly-ferromagnetic phase transition, as observed in  Ce$_x$La$_{1-x}$B$_6$ system \cite{PhysRevLett.82.3669, Harrison2001234}.

On formal side, we would like to note that we included large (but finite) U effects within the Anderson lattice model, with both the hybridization and the Kondo interaction effects incorporated in a systematic manner. In result, the effective Hamiltonian can be termed Anderson-Kondo lattice model. The systematic aspect of our approach to solving the model relies on utilizing the Fukushima
type of approach and generalizing it to a two-orbital situation, in which the constraints ensuring the statistical consistency of the results have been also included. In this manner, we can claim that we have provided a \textit{mutually consistent mean-field approach} starting from the Gutzwiller-type projection for the correlated f-electron states in this two-orbital model. The Appendices \ref{A2} to \ref{A1} provide some of the technical details of our approach.

\begin{figure}
 \includegraphics[width=0.5\textwidth]{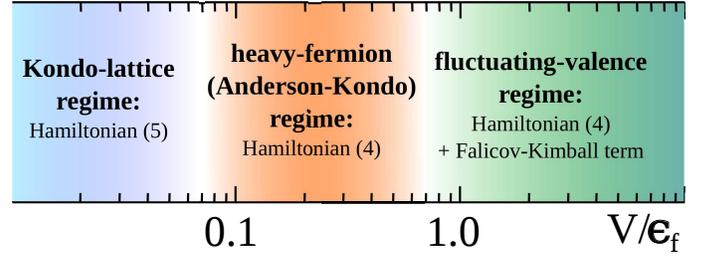}                
  \caption{ (Color online) Schematic representation of various regimes as defined by effective model Hamiltonians. The nature of the borders
  between the regions is not sharply defined. This classification is particularly  relevant for various versions of the so-called renormalized mean-field analysis. }
 \label{fig:fig10}
\end{figure}

Generally, it comes out from our results (as well as from the earlier results \cite{PhysRevB.58.3293,PhysRevB.56.R14239})
that for $|V|< |V_{min}|$ ($\sim 0.25$ for $\epsilon_f = -0.75$) the f-electrons are almost localized ($n_f\rightarrow 1$ c.f. inset, Fig. \ref{fig:mV}; see also Fig. \ref{fig:mfmc}(c)). Well below the critical value $|V_{min}|$ we can say that the Kondo-lattice Hamiltonian (5) is applicable. In the heavy-fermion regime, we have $|V_{min}| \lesssim |V| < |\epsilon_f|$ and the Anderson-Kondo lattice model
in the forms (\ref{eq:3}) or (4) should be used, whereas $V/\epsilon_f \gtrsim 1$ we enter  a fluctuating valence regime,
where the Falicov-Kimball term may also be important to capture the large change in $n_f:n_c$. The discussed subdivision into regimes is illustrated schematically in Fig. \ref{fig:fig10}. Obviously, in all this consideration we assume that $|V|/(U +\epsilon_f)\ll1$.

Finally, the approach is an effective (almost localized) Fermi (non-Landau) picture with renormalized characteristics nonlinear effective fields
($\lambda_f$, $\lambda_{fm}$, $\lambda_{cm}$), spin-dependent effective masses and adjustable chemical potential in each phase. All the parameters are determined in a consistent manner by determining first the effective Gibbs energy functional which plays role of Landau functional with number 
of the mean fields (order parameters). It would be natural to apply the scheme to the analysis of the paired (superconducting) states.
This is planed as the next step. Also, a coexistence of (antiferro)magnetism and superconductivity should be analyzed, since it is observed in quite few  heavy-fermion compounds \cite{PSSB:PSSB200983061, PhysRevLett.95.247004, PhysRevLett.99.146402}.

\begin{acknowledgements}
Enlightening discussions with Drs. Jakub J\c edrak and Jan Kaczmarczyk are gratefully acknowledged.
We wish also  to thank Prof. Wolfgang Nolting from the Humboldt University for a helpful discussion. The work was supported by Ministry of Higher Education and Science, Grants Nos. N
N202 128736 and N N202 489839, as well as by the Project \textit{TEAM} awarded to our group by the Foundation for Polish Science (FNP) for the years 2011-2014.
\end{acknowledgements}

\appendix
\section{Role of the interorbital (Falicov-Kimball) term}
\label{A2}
The Falicov-Kimball term taken in the intraatomic form
\begin{equation}
U_{cf} \sum_{<im>} \hat{N}_i \hat{n}_m = U_{cf} \sum_i\hat{N}_i\hat{n}_i
\end{equation}
enters in two ways in the starting effective Hamiltonian (\ref{eq:3}). First, the denominator of the third term 
should be replaced by $U-U_{cf}+\epsilon_f$, so that the effective Kondo coupling constant increases, since it has then the form $J_{im} = 2|V_{im}|^2/(U+U_{cf}+\epsilon_f)$. Second, after the canonical transformation of (\ref{eq:1}) leading to (\ref{eq:3}) there remains the projected   
part of  repulsive Coulomb interaction term $U_{cf} \sum_{<im>} \hat{\nu}_{if} \hat{n}_m$ which, when treated in the simplest Hartree-Fock manner, reduces to 

\begin{equation}
U_{cf} \sum_{<im>} \hat{\nu}_{if} \hat{n}_m \approx U_{cf} \sum_{<im>}( n_f \hat{n}_m + n_c \hat{\nu}_{if} - n_cn_f).
\end{equation}
In this manner, this term first of all shifts the energy $f$ and $c$ levels. Since such shifts are already included in (\ref{eq8}) via the terms $\sim \lambda _f$ and $\lambda$, respectively, to be calculated self-consistently anyway, we can disregard them. This is particularly so in the interesting us heavy-fermion-liquid regime $\langle \hat{\nu}_{if}\rangle \equiv 1$ ($n_f \rightarrow 1$), and then
\begin{equation}
U_{cf} \sum_{<im>} \hat{\nu}_{if} \hat{n}_m \equiv  U_{cf} \sum_{<im>}(\hat{\nu}_{if}-1)\hat{n}_m + U_{cf} \sum_{<im>} \hat{n}_m.
\end{equation}
 In  effect, the first term is relatively small and hence can be decoupled in the Hartee-Fock manner. Such reasoning justifies qualitatively the fact of neglecting the Falicov-Kimball term in our mean-field-like approach, at least in the heavy-fermion limit.
 
\section{The Fukushima local-constraint scheme of evaluating the expectation values }
\label{A0}
In order to provide an insight into the procedure of calculating the expectation values in the Fukushima local-constraint scheme,
we present here the detailed evaluations of the average representing antiferromagnetic Kondo coupling $\hat{K}\equiv \mathbf{\hat{S}}_i\cdot \mathbf{\hat{s}}_m - \frac{\hat{\nu}_{if} \hat{n}_{m} }{4} + \frac{\hat{\nu}_{if\bar{\sigma}} \hat{n}_{m\sigma} }{4}$ from Eq. (4).
After performing simple algebraic manipulations we can present this term in the following form:
\begin{equation}
\hat{K}_i= -\hat{N}_{if\bar{\sigma}} (1-\hat{N}_{if\sigma}) c^{\dagger}_{m\sigma} c_{m\sigma} - c^{\dagger}_{m\bar{\sigma}} f_{i\bar{\sigma}}
f^{\dagger}_{i\sigma} c_{m\sigma}.
\end{equation}
According to (\ref{eq:6}) we can present expectation value of above operator as  

\begin{eqnarray}
\langle \hat{K}_i \rangle &=&  -\frac{\langle \mathcal{\hat{P}} \hat{N}_{i\bar{\sigma}} (1-\hat{N}_{i\sigma})c^{\dagger}_{m\sigma}c_{m\sigma}\mathcal{\hat{P}}\rangle_0} 
{\langle\mathcal{\hat{P}}^2\rangle_0} \nonumber \\
 &-&\frac{\langle \mathcal{\hat{P}} c^{\dagger}_{m\bar{\sigma}} f_{i\bar{\sigma}}f^{\dagger}_{i\sigma}c_{m\sigma}\mathcal{\hat{P}}\rangle_0} 
{\langle\mathcal{\hat{P}}^2\rangle_0},
\end{eqnarray}
where projector $\hat{\mathcal{P}}$ can be presented in the following form
\begin{eqnarray}
\mathcal{\hat{P}}_i &=& \lambda_{i\uparrow}^{\hat{N}_{i\uparrow}/2} \lambda_{i\downarrow}^{\hat{N}_{i\downarrow}/2}(1-\hat{N}_{i\downarrow}\hat{N}_{i\uparrow} )\nonumber \\
&\equiv& (1-\hat{N}_{i\uparrow})(1-\hat{N}_{i\downarrow}) + \sqrt{\lambda_{i\downarrow}} \hat{N}_{i\downarrow}(1-\hat{N}_{i\uparrow})\nonumber \\
&&+\sqrt{\lambda_{i\uparrow}} \hat{N}_{i\uparrow}(1-\hat{N}_{i\downarrow}).
\end{eqnarray}
The uncorrelated average of the norm is
\begin{equation}
\langle\mathcal{\hat{P}}_i^2\rangle_0 \equiv \frac{(1-n_{if\uparrow})(1-n_{if\downarrow})}{1-n_{if}}.
\end{equation}
Using the Wick theorem we can easily calculate the following averages as
\begin{eqnarray}
&&\langle \mathcal{\hat{P}}_i \hat{N}_{if\bar{\sigma}}(1-\hat{N}_{i\sigma})c^{\dagger}_{m\sigma}c_{m\sigma}\mathcal{\hat{P}}_i\rangle_0\nonumber\\
&=&\langle \lambda_{i\bar{\sigma}}\hat{N}_{i\bar{\sigma}} (1-\hat{N}_{i\sigma})c^{\dagger}_{m\sigma}c_{m\sigma}\rangle_0\nonumber\\
&=& \lambda_{i\bar{\sigma}} \left[ n_{if\bar{\sigma}} (1-n_{if{\sigma}}) n_{ic\sigma} + n_{if\bar{\sigma}} \gamma ^{*}_{im\sigma}\gamma_{im\sigma}\right],
\end{eqnarray}

\begin{eqnarray}
\langle \mathcal{\hat{P}}_i  c^{\dagger}_{m\bar{\sigma}} f_{i\bar{\sigma}}f^{\dagger}_{i\sigma}c_{m\sigma} \mathcal{\hat{P}}_i\rangle_0
&=&\langle \sqrt{\lambda_{i{\sigma}}\lambda_{i\bar{\sigma}}} c^{\dagger}_{m\bar{\sigma}} f_{i\bar{\sigma}}f^{\dagger}_{i\sigma}c_{m\sigma}\rangle_0\nonumber\\
&=& \sqrt{\lambda_{i{\sigma}}\lambda_{i\bar{\sigma}}} \gamma_{im\sigma}\gamma^{*}_{im\bar{\sigma}}. 
\end{eqnarray}
Finally, 
\begin{eqnarray}
\langle \hat{K}_i \rangle &=& - n_{if\bar{\sigma}}n_{ic\sigma} - \frac{n_{if\bar{\sigma}}}{1-n_{if\sigma}}\gamma_{im\sigma}\gamma^{*}_{im\sigma}\nonumber \\ &-&\frac{\gamma_{im\sigma}\gamma^{*}_{im\bar{\sigma}}}{\sqrt{(1-n_{if\sigma})(1-n_{if\bar{\sigma}})}}.
\end{eqnarray}


\section{Density of states and effective-mass of renormalized quasiparticles}
\label{A1}

Using quasiparticle energy spectrum (\ref{eq10}) we can evaluate the density of states by performing the sum  

\begin{equation}
D(\omega) = \frac{1}{\Lambda}\sum_{\textbf{k}\sigma \alpha} \delta(\omega-E_{k\sigma}^\alpha).
\label{eq12}
\end{equation}
This can be carried out numerically by representing delta function as the Gaussian function with a small finite width.
We may also achieve an analytical form for DOS, as follows.
The matrix of the coefficients (\ref{eqa1}) from the equation (\ref{eq9}) can be shifted \cite{2005condmat} so that obtained matrix
\begin{equation}
\mathcal{K}_\lambda \equiv 
\begin{pmatrix}
\epsilon^c_{\textbf{k} \sigma} & -\tau_{\sigma}\\
-\tau_{\sigma} & \epsilon^f_{\sigma}\\
\end{pmatrix},
\label{eqa1}
\end{equation}
has a triangular form
\begin{equation}
\mathcal{K}_\lambda + \left(\lambda + \lambda_f + \lambda_{fm} \sigma + \mu \right)\mathbb{I}  = 
\begin{pmatrix}
\xi_{\textbf{k}} - b & -\tau_{\sigma}\\
-\tau_{\sigma} & 0\\
\end{pmatrix},
\end{equation}
where $b = -\lambda_f -\sigma(\lambda_{fm}-\lambda_{cm})$ and $\xi_{\textbf{k}} = -2\eta (\cos{k_x} +\cos{k_y}) $.
The eigenvalues of the above matrix are as follows
 \begin{equation}
e_{\textbf{k}\sigma}^n = \frac{1}{2}\left(\xi_{k}-b + n\sqrt{(\xi_{k} -b)^2 + 4\tau_{\sigma}^2}\right). 
\end{equation}
Then, it easy to see that the sum
\begin{equation}
D_e(\omega) = \frac{1}{\Lambda}\sum_{\textbf{k}\sigma n } \delta(\omega - e_{k\sigma}^n )
\label{eqa3}
\end{equation}
corresponds to the true DOS (\ref{eq12}) shifted by the value $\lambda + \lambda_f + \lambda_{fm} \sigma + \mu$   
To simplify the summation  (\ref{eqa3}) we use properties of delta function 
so that $D_e(\omega)$ can be rewritten in the form

\begin{equation}
D_e(\omega) = \frac{1}{\Lambda}\sum_{\textbf{k}\sigma}\sqrt{(\xi_{\textbf{k}} -b)^2 + 4\tau_{\sigma}^2} \  \delta(\omega^2 -\omega(\xi_{\textbf{k}}-b) - \tau_{\sigma}^2).
\end{equation}
In the thermodynamic limit the energy spectrum is extremely dense so that it can be presented as smooth function and summation can be substituted by integral so that $e_{\textbf{k}\sigma}^n \rightarrow \omega$. Using this we can express
 $(\xi_{\textbf{k}}-b) \rightarrow \frac{\omega^2 - \tau_{\sigma}^2}{\omega}$ so that

\begin{equation}
D_e(\omega) = \frac{1}{\Lambda}\sum_{\textbf{k}\sigma}\left(1+\frac{\tau_{\sigma}^2}{\omega^2}\right)\delta\left(\xi_{\textbf{k}} - b -\omega\left(1-\frac{\tau_{\sigma}^2}{\omega^2}\right)\right).
\label{eqa5}
\end{equation}
For the noninteracting electrons density of states can be written
as the product of a line-shape function and a heaviside
function, which ensures that the density of states
vanishes outside the band: $D(\omega) = g(\omega)\Theta(W^2 - 4\omega^2)$, where $W$ is the width of a band.
In the similar way manner Eq. (\ref{eqa5}) can be expressed in the form:

\begin{equation}
D_e(\omega) = \frac{1}{\Lambda}\sum_{\sigma}\left(1+\frac{\tau_{\sigma}^2}{\omega^2}\right)g(f(\omega))\Theta\left(W^2 - 4 f(\omega)^2\right),
\end{equation}
where $f(\omega) = b + \omega\left(1-\frac{V^2}{\omega^2}\right)$.
The edges of DOS for interacting DOS are given by the solution of equation $f(\omega)=0$ and are as follows
\begin{eqnarray}
\omega_{1\sigma} &=& -\sqrt{(b/2 + W/2)^2+ \tau_{\sigma}^2}-W/4 -b/2, \nonumber\\
\omega_{2\sigma} &=& -\sqrt{(b/2 - W/2)^2+ \tau_{\sigma}^2}+W/4 -b/2, \nonumber\\
\omega_{3\sigma} &=& \sqrt{(b/2 + W/2)^2+ \tau_{\sigma}^2}-W/4 -b/2, \nonumber\\
\omega_{4\sigma} &=& \sqrt{(b/2 - W/2)^2+ \tau_{\sigma}^2}+W/4 -b/2.
\end{eqnarray}

The DOS has non-zero values only in two intervals $(\omega_{1\sigma},\omega_{2\sigma})$ and $(\omega_{3\sigma},\omega_{4\sigma})$. 
The resulting expression for $D_e(\omega)$ is finally as follows:
\begin{equation}
D_e(\omega) = \frac{1}{W}\sum_{\sigma i}\left(1+\frac{\tau_{\sigma}^2}{\omega^2}\right)\Theta\left(\omega - \omega_{i\sigma}\right)(-1)^{i+1}.
\label{eqa9}
\end{equation}
 
We define the effective mass enhancement of quasiparticles  in the standard manner 

\begin{equation}
m_\sigma ^* \equiv \frac{m_\sigma ^*}{m} = \left(\frac{\partial E^n_{k\sigma}}{\partial \epsilon_{k\sigma}^c}\right)^{-1}\Biggr|_{\mu}.
\end{equation} 
Using the  same substitution as above (i.e., $E^n_{k\sigma} \rightarrow \omega$), we can 
express the starting  energy of free electrons by $\epsilon_{k\sigma}^c \rightarrow \frac{\omega^2 - \tau_{\sigma}^2 -\omega \epsilon_\sigma^f}{\omega -\epsilon^f_\sigma}$, so that finally the mass enhancement at the Fermi energy is

\begin{equation}
m_\sigma ^* =\frac{\partial}{\partial \omega} \left(\frac{\omega^2 - \tau_{\sigma}^2 -\omega \epsilon_\sigma^f}{\omega -\epsilon^f_\sigma}\right) =   1+ \frac{\tau_\sigma^2}{(\epsilon^f_\sigma - \omega)^2}\Biggr|_{\omega = 0}.
\end{equation}
This expression is used for numerical calculations in Sec. \ref{results}.    

\bibliography{bib}

\end{document}